\documentclass[12pt]{article}
\usepackage{epsfig}
\usepackage{amsmath}
\usepackage{amssymb}
\usepackage{amsfonts}
\usepackage{enumerate}
\usepackage{cite}
\newcommand{\Hrule}{\vskip-10mm\begin{flushleft}\rule{\linewidth}{0.1mm}\end{flu
shleft}\vskip-1.5mm}
\title{An alternate spintronic analog of the electro-optic modulator}
\author{S. 
Bandyopadhyay$\thanks{Corresponding author. E-mail: sbandy@vcu.edu}$ \\ 
 Department of Electrical Engineering\\
Virginia Commonwealth University, Richmond, Virginia 23284\\
\\M. Cahay \\
Department of Electrical and Computer Engineering and Computer 
Science\\
University of Cincinnati, Cincinnati, Ohio 45221}
\date{}
\begin{document}
\maketitle

\baselineskip=24pt

There is significant current interest in spintronic devices fashioned after a 
spin analog of the electro-optic modulator proposed by Datta and Das [Appl. 
Phys. Lett., \underline{56}, 665 (1990)]. In their modulator, the ``modulation'' 
of the spin polarized current
is carried out by tuning the Rashba spin-orbit interaction with a gate voltage. 
Here, we propose an analogous modulator where the modulation is carried out by 
tuning the Dresselhaus spin orbit interaction instead. The advantage of the 
latter is that there is no magnetic field in the channel unlike in the case of 
the Datta-Das device. This can considerably enhance modulator performance.

\bigskip

\noindent PACS: 85.75.Hh, 85.35.Be, 72.25.Dc

\pagebreak

In 1990, Datta and Das proposed a spintronic analog of the electro-optic 
modulator \cite{datta}. It consists  of a quasi 
one-dimensional 
semiconductor 
channel with ferromagnetic source and drain contacts (Fig. 1(a)). Electrons are 
injected with a definite spin orientation from the source,  
which is then controllably precessed in the channel with a gate-controlled 
Rashba spin-orbit
interaction \cite{rashba}, and finally sensed 
at the drain. At the drain end, the electron's transmission probability depends 
on the relative 
alignment of its spin with  the drain's (fixed) magnetization.  By controlling 
the angle of spin precession in the channel with a gate voltage, 
one can control the relative spin alignment at the drain end, and hence control 
the 
source-to-drain  current. This realizes the basic ``transistor'' action. Because 
of this attribute, the Datta-Das device came to be known as the ballistic Spin 
Field Effect Transistor (SPINFET).

Despite the fact that the SPINFET was proposed more than a decade ago, it has 
never been experimentally realized. Recently, we found that one of the serious 
impediments to its realization is the presence of a magnetic 
field in its channel  caused by the ferromagnetic source and drain contacts.
This field has been ignored in practically all past work, but  has crucial 
consequences. Based on available data for  device configurations that are 
similar to the SPINFET \cite{wrobel}, we estimate that in a 0.2 $\mu$m long
channel, the average magnetic field may approach 1 Tesla. This field  has many 
deleterious effects \cite{cahay_prb1, cahay_prb2}. First, it results in  a 
Zeeman spin splitting that affects the dispersion relations of the Rashba spin 
split subbands in the channel. Consequently, there is
``spin mixing'' in each subband, so that no subband has a definite spin 
quantization axis 
\cite{cahay_prb1}. As a result,  non-magnetic 
scatterers can flip spin \cite{cahay_prb2} thereby  making  spin transport 
non-ballistic in the presence of 
 normal impurities, surface roughness, etc., which otherwise would not have 
 affected spin transport. Second, the ``phase shift'' of the spintronic 
modulator
will be no longer independent of energy \cite{cahay_prb1, cahay_prb2} (in ref. 
1, it was claimed to be independent of energy because the channel magnetic field 
was ignored). Therefore, 
 ensemble averaging over electron energy will dilute the modulation effect.  
Suffice it to say then that it is important to eliminate the magnetic field in 
the channel.

Although it is possible to engineer the Datta-Das device to reduce the channel 
field, it can never be completely eliminated (unless complicated spin filter 
devices \cite{koga} are employed) since the magnetization in the source and 
drain contacts have to be always {\it along} the channel. The only other 
solution is to find an alternate analogous device where the magnetic fields due 
to the source and drain contacts are {\it transverse} to the channel. Here, we 
do precisely that and propose an alternate device, based on the Dresselhaus spin 
orbit interaction \cite{dresselhaus} rather than the Rashba interaction. In this 
device, the source/drain 
magnetization will be {\it transverse} to the channel, which vastly reduces the 
channel magnetic field. The only channel field that could be present is the 
fringing field at the edges adjoining the source and drain contacts. This is 
negligible.

Our device is schematically shown in Fig 1(b) and 1(c). Since it has no {\it 
structural} inversion asymmetry, we can ignore the Rashba interaction. However, 
there is a bulk inversion asymmetry in the channel material that ensures the 
presence of a Dresselhaus interaction. We will also assume a strictly 
one-dimensional (1-d) channel (only the lowest subband is occupied by carriers) 
in order to extract the best device performance. The need for one dimensionality 
was already elucidated in ref. 1. Furthermore, since there is no Dyakonov-Perel' 
spin relaxation in a strictly 1-d channel {\it in the absence of a channel 
magnetic field} \cite{cond-mat}, we can expect nearly ballistic spin transport. 
Following usual procedure, the 1-d channel will be defined by split gates 
\cite{vanwees} on the surface of a 
quantum well heterostructure.

The single-particle Hamiltonian describing an electron in the 1-d channel of 
this device is
\begin{equation}
H = \epsilon + {{\hbar^2 k_x^2}\over{2 m^*}} + 2 a_{42} \sigma_x k_x \left [
{{m^* \omega}\over{2 \hbar}} - \left ( {{\pi}\over{W_y}} \right )^2 \right ]
\end{equation}
where $\epsilon$ is the lowest subband energy, $a_{42}$ is the material constant 
associated with the strength of the Dresselhaus interaction \cite{cardona}, 
$\sigma$ is the Pauli spin matrix, and $W_y$ is the channel dimension in the 
y-direction. We assume the potential profile in the y-direction to be a square 
well with hardwall boundaries and the potential profile in the z-direction is
parabolic since confinement in this direction is enforced by split gates. The 
curvature of the parabolic potential is $\omega$ which can be tuned by 
varying the applied voltage on the Schottky split gates. Here, we have assumed a 
direct gap semiconductor. The Dresselhaus spin orbit interaction term has a 
subtle dependence on the crystallographic orientation of the channel 
\cite{dietl}, but it is not {\it qualitatively} important in the present 
context. It may however assume importance in device optimization.

The rest of the analysis is fashioned after ref. 1. Diagonalizing the 
Hamiltonian in Equation (1), we find that the eigenspinors in the channel are 
[1~1]$^{\dag}$ and [1~-1]$^{\dag}$ which are +x-polarized and -x-polarized 
states. They have eigenenergies that differ by 2$\beta k_x$ where $\beta$ = 2 
$a_{42} [
m^* \omega/(2 \hbar) -  ( \pi/W_y )^2 ]$. Accordingly,
\begin{eqnarray}
E(+x~pol.) & = & \epsilon + \hbar^2 k_{x+}^2/2m^* + \beta k_{x+} \nonumber \\
E(-x~pol.) & = & \epsilon + \hbar^2 k_{x-}^2/2m^* -\beta k_{x-}
\label{energies}
\end{eqnarray}

An electron incident on the channel with energy $E$ will have 
two different wavevectors  $k_{x+}$ or $k_{x-}$ depending on whether 
its spin is +x or -x-polarized. Now, if we inject a +z-polarized electron
into the channel from the source contact, it will couple equally to the +x and 
-x-polarized 
subbands since
\begin{eqnarray}
\left [ \begin{array}{c}
             1\\
             0\\
             \end{array}   \right ] 
 =            
 \left [ \begin{array}{c}
1\\
1 \\ 
\end{array} \right ] +
\left [ \begin{array}{c}
             1\\
             -1\\
             \end{array}   \right ]  
\end{eqnarray}

At the drain end, the eigenspinor will be $[e^{i k_{x+} L} + e^{i k_{x-} L} ~~
e^{i k_{x+} L} - e^{i k_{x-} L}]^{\dag}$, where $L$ is the channel length. If 
the drain is magnetized in the 
+z direction, then the transmission probability (and therefore the source to 
drain current) will be proportional to $|[1~~0] [e^{i k_{x+} L} + e^{i k_{x-} L} 
~~
e^{i k_{x+} L} - e^{i k_{x-} L}]^{\dag}|^2$ = 4 $cos^2 [(k_{x-} - k_{x+})L/2]$
= 4 $cos^2[m* \beta L/\hbar^2]$, where we have used Equation (\ref{energies}) to 
arrive at the last equality.

It is obvious now that this device is an exact analog of the device in ref. 1. 
As in ref. 1, we point out that the phase shift between the two orthogonal spin   
states (+x and -x polarized) is $\Delta \phi$ (= $2m* \beta L/\hbar^2$) which is 
independent of the electron wavevector (or energy). Therefore the interference 
between the two spin states causing the conductance modulation survives 
ensemble averaging over the electron energy at elevated temperatures. Actually, 
this is only strictly true if there is no channel magnetic field 
\cite{cahay_prb1, cahay_prb2}. In the Datta-Das device, this would {\it not} 
have been strictly true because of the channel magnetic field, but in our case, 
it is.

The crucial difference between this device and that in ref. 1 is that here the 
contacts have to be magnetized in the z-direction so that the 
magnetic field caused by the contacts is {\it perpendicular} to the channel 
which is in the x-direction. That is why, we can neglect any Zeeman spin 
splitting in the channel which we could not do for the device in ref. 1. As 
mentioned before,
this Zeeman spin splitting (or the channel magnetic field) would have been 
harmful to the device in many ways. 

Before concluding, we can compare the minimum channel lengths $L_{min}$ required 
to  cause a phase shift of $\pi$ radians between the two spin states. The 
channel must be at least this long in order to observe one complete cycle of
switching from the maximum to the minimum conductance state. Comparing  the 
two devices:
\begin{equation}
{{L_{min}|ref.~1}\over{L_{min}|this~device}} = {{\beta}\over{\eta}} \approx {{
a_{42}m^* \omega/(2 \hbar)}\over{a_{46} {\cal E}}}
\end{equation}
where $\eta$ is the strength of Rashba coupling as defined in ref. 1, $a_{46}$ 
is a material 
parameter indicative of the degree of Rashba coupling and ${\cal E}$ is the 
interface electric field causing the Rashba coupling. In GaAs, $a_{42}$ is 
calculated to be 2.9$\times$10$^{-29}$ eV-m$^3$ \cite{cardona}, $a_{46}$ is 
calculated as 
9$\times$10$^{-39}$ C-m$^2$ \cite{lommer}, and ${\cal E}$ can be as high as 
300 kV/cm. We will assume that $\hbar \omega$ = 25 meV ($\hbar \omega$ $\approx$ 
 25 meV was achieved in ref. \cite{vanwees}). Based on these figures, 
$L_{min}|this~device$ = 0.36$L_{min}|ref.~1$, so that the two lengths are 
comparable (of the same order). 

In conclusion, we have proposed a device which is analogous to the spintronic 
modulator proposed in ref. 1, but has the additional advantage of being immune
to spin mixing effects in the channel, spin flip by non-magnetic scatterers,
and dilution of the modulation by ensemble averaging over the electron 
energy. All this has been achieved by eliminating the channel magnetic field. 
The fabrication of this device is no more difficult than fabricating the 1-d 
SPINFET of ref. 1; in fact, it may be somewhat simpler since we do not 
need a top gate (or back gate) to induce the Rashba effect. It is possible that 
this device may be easier to implement, and may be somewhat more robust than the 
device of ref. 1.

\pagebreak

\pagebreak

\noindent {\bf Figure captions}

\noindent {\bf Fig. 1}: (a) Schematic of the spintronic modulator of ref. 1.
(b) side view of the spintronic modulator proposed in this work, (b) top view 
showing the split gates.

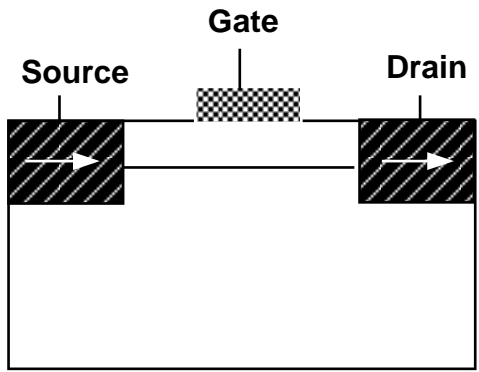
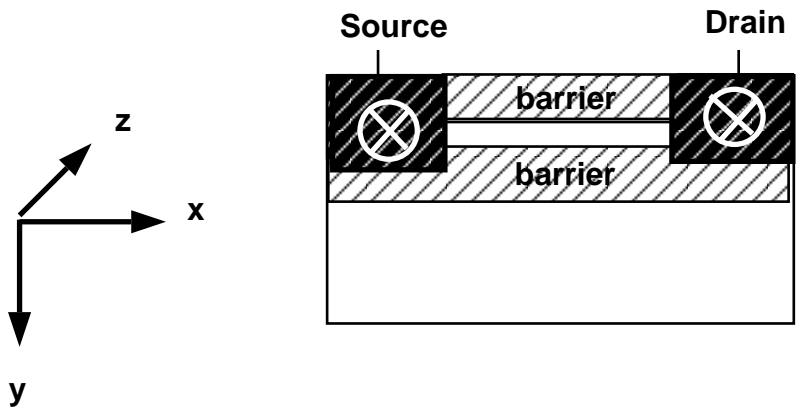
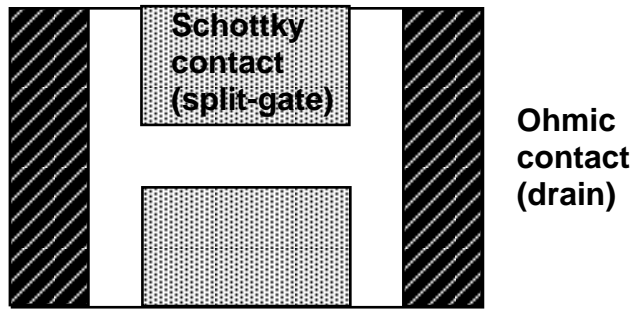
\end{document}